\documentclass[pre,amsmath,aps,superscriptaddress,twocolumn]{revtex4}
\usepackage{graphicx,xspace}
\usepackage{float}
\usepackage{amsmath}
\usepackage{amssymb}
\usepackage{color,soul}
\usepackage{epsfig}
\usepackage{graphicx,psfrag,xspace}

\begin{document}

\title{A continuous time random walk model of transport in variably saturated heterogeneous porous media}
\author{Andrea Zoia}
\email{andrea.zoia@cea.fr}
\affiliation{CEA/Saclay, DEN/DM2S/SFME/LSET, B\^at.~454, 91191 Gif-sur-Yvette Cedex, France}
\author{Marie-Christine N\'eel}
\affiliation{Universit\'e d'Avignon et des Pays de Vaucluse, UMR 1114 EMMAH, 84018 Avignon Cedex, France}
\author{Andrea Cortis}
\affiliation{Earth Sciences Division, Lawrence Berkeley National Laboratory, Berkeley, California 94720, USA}

\begin{abstract}
We propose a unified physical framework for transport in variably saturated porous media. This approach allows fluid flow and solute migration to be treated as ensemble averages of fluid and solute particles, respectively. We consider the cases of homogeneous and heterogeneous porous materials. Within a fractal mobile-immobile (MIM) continuous time random walk framework, the heterogeneity will be characterized by algebraically decaying particle retention-times. We derive the corresponding (nonlinear) continuum limit partial differential equations and we compare their solutions to Monte Carlo simulation results. The proposed methodology is fairly general and can be used to track fluid and solutes particles trajectories, for a variety of initial and boundary conditions.
\end{abstract}
\maketitle

\section{Introduction}

Accurately predicting the spreading of a chemical species through unsaturated porous materials (i.e., materials with locally-varying fluid content) is key to mastering such technological challenges as polluted sites remediation, environmental protection, and waste management~\cite{muskat, hilfer, bear, cortis_flow, sahimi, demarsily}. The evolution of the solute concentration profile in the traversed medium cannot be a priori decoupled from that of the fluid flow, which is ultimately responsible for the advection and dispersion mechanisms of the solutes. Moreover, the effects of the fluid distribution are often combined with those of spatial heterogeneities. Such heterogeneities can be present both at the pore-scale (microscopic) and at the Darcy-scale (macroscopic). As a consequence, experimental results reported in literature often show that solutes concentration displays non-Fickian features, such as breakthrough curves with long tails and non-Gaussian spatial profiles~\cite{bromly, cortis_flow, levy}.

In this respect, there exists an increasing need for reliable numerical techniques to tackle flow and transport problems. In this work, we address the issue of determining the fluid content and the solute concentration profiles within unsaturated homogeneous as well as heterogeneous media by resorting to a random walk approach. Random walks are extensively used to describe solutes transport in saturated media~\cite{Madziarz, delay}, although their application to unsaturated flows appears to be somehow neglected~\cite{buckergittel}. While complementing each other, the random walk and the continuum-limit approaches display specific advantages and disadvantages. Random walks, for instance, do not introduce the spurious numerical dispersion typical of Eulerian (continuum) numerical schemes~\cite{Madziarz, delay, buckergittel}. As such, random walks are particularly well suited to deal with unsaturated materials, where sharp contrasts between stagnant and fluid-saturated regions, or at macroscopic heterogeneity interfaces, may give rise to steep propagating fronts. Eulerian (continuum) numerical schemes, on the other hand, are generally faster than the corresponding Monte Carlo simulations.

We begin our analysis by illustrating flow and transport in homogeneous media, and detail how the Richards equation for the fluid flow and the Advection-Dispersion Equation (ADE) for the solutes can be recast in a Fokker-Planck Equation (FPE) form. The FPE governs the probability density (pdf) of finding a walker (a fluid or solute parcel, respectively) at a given position at a given time. The central idea is that these walkers perform stochastic trajectories in the traversed medium. Taking the ensemble average of the fluid and solute parcels trajectories yields the desired macroscopic quantities, i.e., the fluid content and solutes concentration profiles. As the fluid movement and the solute transport are (nonlinearly) coupled via the macroscopic governing equations, the underlying stochastic trajectories also display a nonlinear coupling. On the other hand, the homogeneity hypothesis ensures that the trajectories carry no memory of the past, so that the particles dynamics is Markovian \cite{bhattacharya}.

Then, we focus our attention on unsaturated heterogeneous materials, where non-Fickian behaviors are enhanced by the interplay between nonlinearities in the flow patterns and complex spatial structures: the relative strength of these processes determines the precise details of the solutes distribution. We model the effects of complex nonhomogeneous spatial structures by introducing the possibility of trapping events between successive particles displacements, as customary within a continuous time random walk (CTRW) approach~\cite{rev_geo, lax}. The resulting broad distribution of waiting times at each visited site characterizes the broad velocity spectrum that is often observed in heterogeneous media.

This paper is organized as follows: in Sec.~\ref{physics}, we revise the physical equations that govern the coupled flow-transport problem for nonstationary variable-saturation conditions. Then, in Sec.~\ref{random walk} we present a general nonlinear random walk approach to the simulation of fluid and contaminant particles in locally homogeneous media. These simulation schemes are compared with numerical solutions of the governing equations in Sec.~\ref{sim_homogeneous} for a variety of initial and boundary conditions. The case of discontinuous physical properties of the traversed media is addressed, as well. In Sec.~\ref{heterogeneous} we extend our results to heterogeneous porous media, and derive the corresponding macroscopic governing equations. Finally, conclusions are drawn in Sec.~\ref{conclusions}.

\begin{figure}[t]
\centerline{\epsfxsize=9.0cm\epsfbox{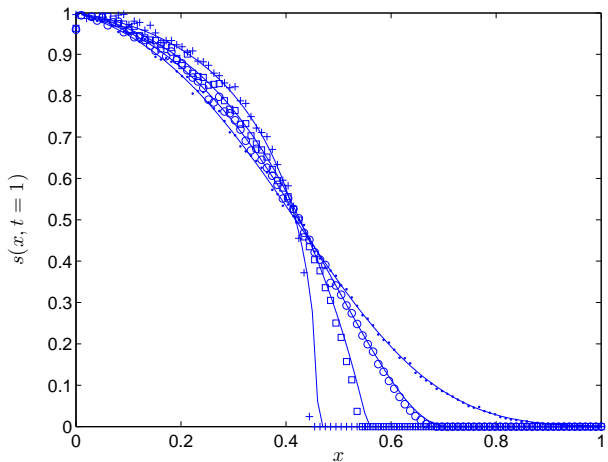}}
\caption{Saturation level spatial profiles $s(x,t)$ at fixed time $t=1$, for varying $\chi$. Numerical integration of Eq.~\eqref{ade_S} is displayed as solid lines, Monte Carlo simulation as symbols: $\chi=0.1$ (dots), $\chi=0.5$ (circles), $\chi=1$ (squares), and $\chi=2$ (crosses). The other coefficients are: $\kappa_0=0.05$, $v_0=0.3$, and $\nu=0.4$.}
   \label{fig1}
\end{figure}

\section{Governing equations in unsaturated homogeneous media}
\label{physics}

In the following, we shall briefly recall the physical equations that govern non-stationary flow-transport processes in unsaturated homogeneous porous media. Consider a vertical column of length $\ell$ and radius $r \ll \ell$, so that the flow-transport process can be considered one-dimensional along the longitudinal direction. Let the column be filled with a homogeneous porous material, and suppose that the medium has an initial variable saturation. The fluid flow dynamics within such region can be described in terms of the (dimensionless) volumetric fluid content $0<s(x,t)<1$~\cite{bear, bhattacharya}, whose evolution is ruled by the continuity equation
\begin{equation}
\partial_t s(x,t) = - \partial_x j_s(x,t),
\label{eq:cont}
\end{equation}
provided that the porosity is constant, i.e., the soil skeleton is rigid~\cite{gerolymatou}. When $s(x,t)=1$ everywhere, the medium is fully saturated in fluid. The so-called Buckingham-Darcy flux (or generalized Darcy's law) $j_s(x,t)$ [L/T] is provided by the constitutive equation
\begin{equation}
j_s(x,t)=K(s)\left[1-\partial_x h(s) \right],
\label{flux_S}
\end{equation}
where the quantity $K(s)$ [L/T] is the saturation-dependent hydraulic conductivity, and $h(s)$ [L] is the saturation-dependent capillary pressure~\cite{bear, bhattacharya}. Then, by introducing the capillary diffusivity $\kappa(s)=K(s){\partial h}/{\partial s}$ [L$^2$/T], we can combine Eqs.~\eqref{eq:cont} and~\eqref{flux_S} so to obtain
\begin{figure}[t]
\centerline{\epsfxsize=9.0cm\epsfbox{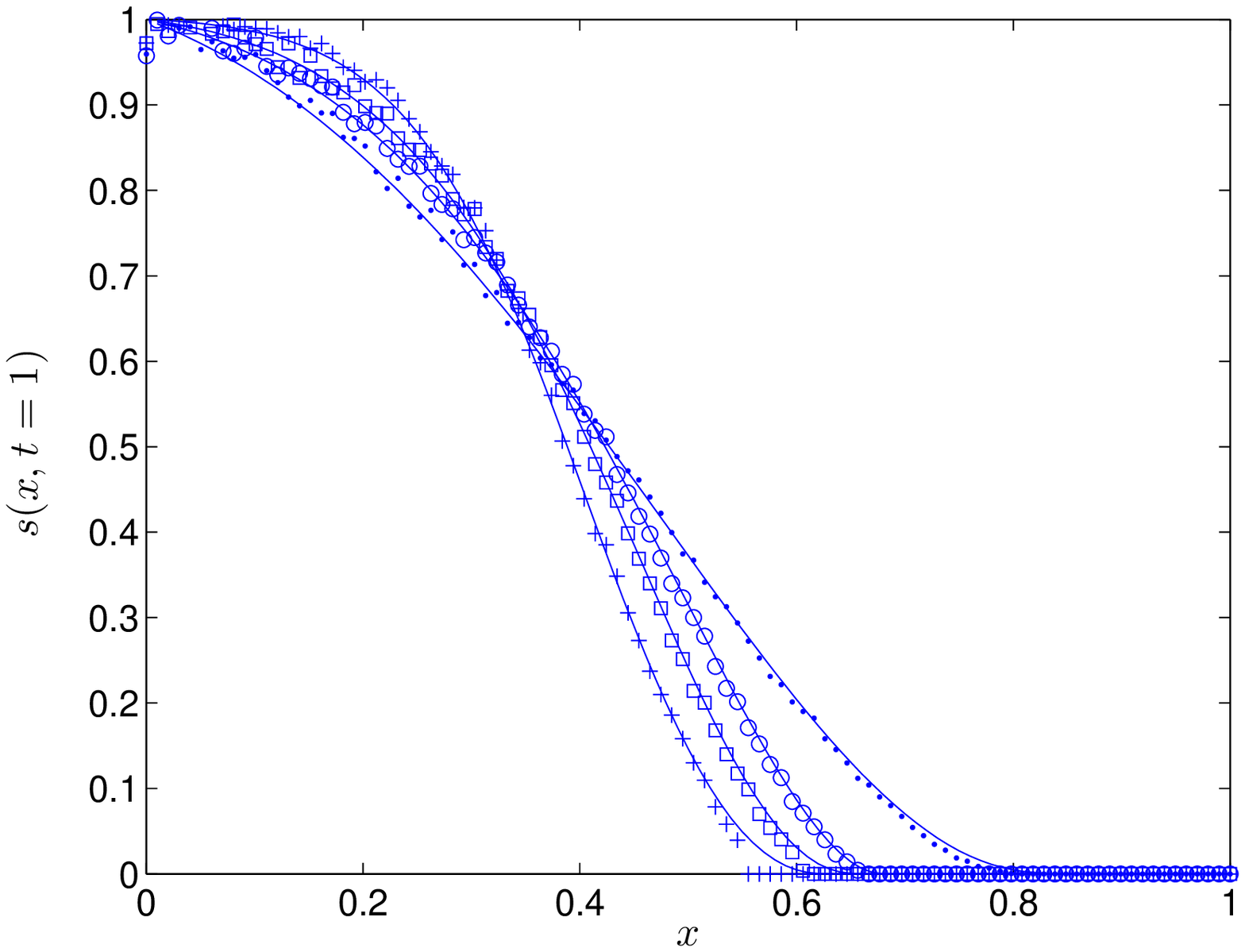}}
\caption{Saturation level spatial profiles $s(x,t)$ at fixed time $t=1$, for varying $\nu$. Numerical integration of Eq.~\eqref{ade_S} is displayed as solid lines, Monte Carlo simulation as symbols: $\nu=0$ (dots), $\nu=0.5$ (circles), $\nu=1$ (squares), and $\nu=2$ (crosses). The other coefficients are: $\kappa_0=0.05$, $v_0=0.3$, and $\chi=0.5$.}
   \label{fig2}
\end{figure}
\begin{equation}
\partial_t s(x,t) = \partial_x \kappa(s) \partial_x s(x,t) - \partial_x K(s).
\label{richards}
\end{equation}
In this form, Eq.~\eqref{richards} has been first derived by Richards~\cite{rich_orig}. The Richards equation has been extensively adopted in describing the dynamics of fluid flows during wetting processes in soils (see the discussion in~\cite{gerolymatou} and references therein). Without loss of generality, we can finally put this equation in conservative form by conveniently defining a `velocity' $v(s)=K(s)/s$ [L/T]:
\begin{equation}
\partial_t s(x,t) = - \partial_x [ v(s) - \kappa(s) \partial_x ] s(x,t).
\label{ade_S}
\end{equation}
The term $v(s)$ plays the role of an effective velocity for the fluid particles and represents the gravitational contribution to the flow dynamics. Remark that equation~\eqref{ade_S} is nonlinear, in that the diffusion $\kappa(s)$ and advection $v(s)$ coefficients depend in general on $s$. In some special cases, it is nonetheless possible to obtain analytical solutions, by resorting to the scaled (Boltzmann) variable $\epsilon=x t^{-1/2}$~\cite{pachepsky, gerolymatou}.

Flow dynamics must be supplemented by the initial and boundary conditions. To set the ideas, as a representative example we may impose $s(0,t)=1$, i.e., we keep the inlet on the column at a constant full saturation. This condition may be physically achieved by putting the porous column in contact with a fluid reservoir (infiltration process). Along the column, we initially assign a given saturation distribution, for instance a constant profile $s(x,0)=s_0$, for $x>0$. Finally, at the outlet of the column we prescribe a vanishing diffusive flux, $\partial_x s(x,t) \vert_{x=\ell}=0$, (Neumann boundary condition), i.e., a flat concentration profile.

\begin{figure}[t]
\centerline{\epsfxsize=9.0cm\epsfbox{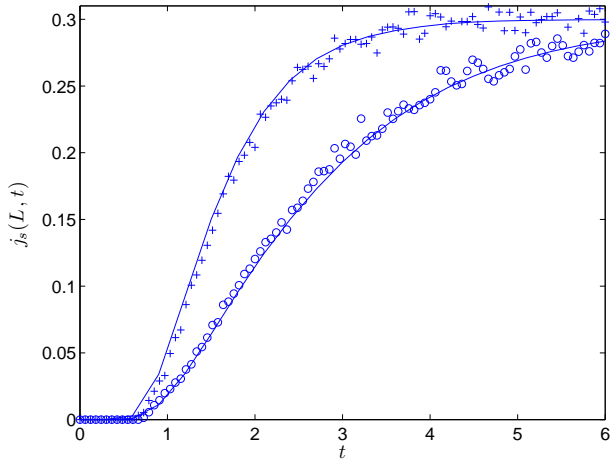}}
\caption{Breakthrough curves $j_s(\ell,t)$ as a function of time $t$ for discontinuous $\kappa$ coefficient. The interface is located at $x_d=\ell/2$. Numerical integration is displayed as solid lines, Monte Carlo simulation as symbols: $\kappa_0=0.4$ in the left layer and $\kappa_0=0.05$ in the right layer (crosses); $\kappa_0=0.05$ in the right layer and $\kappa_0=0.4$ in the left layer (circles). The other coefficients are: $\nu=0.2$, $v_0=0.3$, and $\chi=0.1$.}
   \label{fig3}
\end{figure}

We assume now that a (non-reactive) tracer, e.g., some chemical species, flows diluted in the fluid which is injected into the porous column. Provided that the medium is sufficiently homogeneous, and that physical-chemical interactions of the transported species with the porous matrix and preferential flows can be excluded~\cite{cortis_homog, park}, the solutes dynamics obeys an ADE with saturation-dependent coefficients
\begin{equation}
\partial_t s c(x,t)= - \partial_x [ u(s) - s D(s) \partial_x ] c(x,t),
\label{ade_C}
\end{equation}
where $c(x,t)$ is the solutes concentration, and $s=s(x,t)$.

The advection term $u(s)$ is determined by the fluid flow, namely $u(s)=j_s(x,t)$, whereas the effective dispersion coefficient $D(s)$ accounts for the effects of mechanical dispersion and molecular diffusion mechanisms~\cite{bear}. Note that Eq.~\eqref{ade_C} is linear, although knowledge of the saturation $s(x,t)$ is required in order to determine $c(x,t)$, i.e., problems~\eqref{ade_C} and~\eqref{ade_S} are inherently coupled. Owing to this coupling and to the nonlinearities, the flow patterns are not homogeneous, so that the evolution of the solutes dynamics usually displays non-Fickian features, such as long tails and non-Gaussian shapes. We will discuss this point in detail in the next Section. When the saturation level is uniform within the column, i.e., $s(x,t)=s_0$, Eq.~\eqref{ade_C} reduces to a standard ADE with constant coefficients and contaminant transport becomes Fickian.

\begin{figure}[t]
\centerline{\epsfxsize=9.0cm\epsfbox{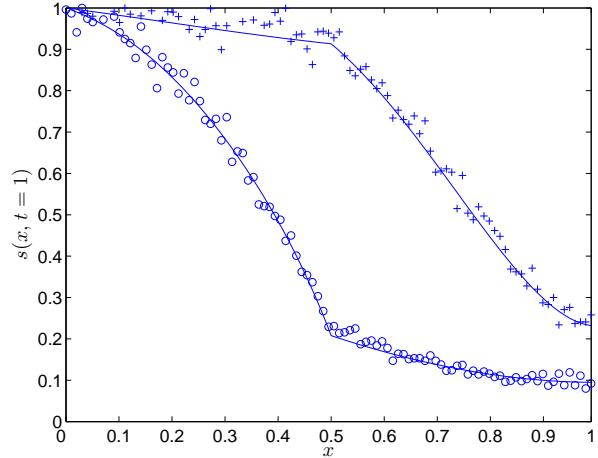}}
\caption{Saturation level spatial profiles $s(x,t)$ at fixed time $t=1$. Numerical integration is displayed as solid lines, Monte Carlo simulation as symbols: $\kappa_0=0.4$ in the left layer and $\kappa_0=0.05$ in the right layer (crosses); $\kappa_0=0.05$ in the right layer and $\kappa_0=0.4$ in the left layer (circles). The other coefficients are: $\nu=0.2$, $v_0=0.3$, and $\chi=0.1$.}
   \label{fig4}
\end{figure}

Concerning initial and boundary conditions for the solute species, in the following we will assume that contaminant release occurs within a given time interval $t_0 \le t \le t_c$, and that during this time span the pollutants concentration at the column inlet has a constant value $c_0$, i.e., $c(0,t_0 \le t \le t_c)=c_0$. Such finite-extension contaminant spills are commonly encountered in environmental remediation problems~\cite{bear}. Before injection, there is no contaminant within the column, i.e., $c(x,t \le t_0)=0$. The boundary condition for the concentration at the outlet is dictated by the outlet boundary for the flow, i.e., it must be a Neumann boundary condition, $\partial_x c(x,t) \vert_{x=\ell}=0$.

\section{A random walk approach}
\label{random walk}

Flow-transport equations \eqref{ade_S} and~\eqref{ade_C} share a similar structure, and can be both written in conservative form as
\begin{equation}
\partial_t \theta p(x,t)= -\partial_x \left[\theta q - \theta d \partial_x \right]p(x,t),
\label{eq_ade_gen}
\end{equation}
for the evolution of the field $p(x,t)$. The coefficients $\theta=\theta(x,t)$, $q=q(x,t)$, and $d=d(x,t)$ are generally space and time dependent. This conceptual picture allows Eqs.~\eqref{ade_S} and~\eqref{ade_C} to be expediently solved by resorting to a random walk formulation. The key idea is to think of the evolving (fluid or contaminant) plume, whose dynamics is described by Eq.~\eqref{eq_ade_gen}, as being composed of a large number of particles performing stochastic trajectories in the traversed porous medium. Then, the quantity $p(x,t) \ge 0$ can be given a probabilistic interpretation (up to a normalization factor), i.e., $p(x,t)$ represents the probability density of finding a walker at a given position $x$, at time $t$.

\begin{figure}[t]
\centerline{\epsfxsize=9.0cm\epsfbox{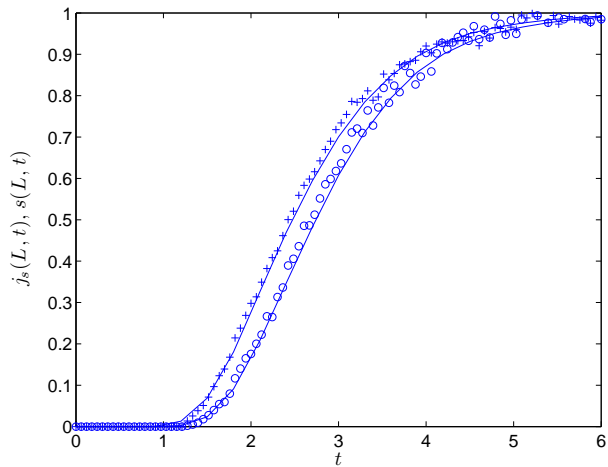}}
\caption{Breaktrough curve (circles) and saturation level (crosses) at the column outlet, normalized so to have unit maximum. Solid lines correspond to numerical integration, symbols to Monte Carlo simulation. A discrepancy between the profiles is apparent for non-constant velocities $v$. The coefficients are: $\kappa_0=0.05$, $\nu=0.4$, $v_0=0.3$, and $\chi=0.1$.}
   \label{fig5}
\end{figure}

Consider an ensemble of $N$ particles at positions $x_j(t)$, $j=1,...,N$. Assume that the stochastic dynamics of each walker is governed by a Langevin-type equation
\begin{equation}
x_j(t+\tau)=x_j(t)+ A(x,t) \tau + \sqrt{2 B(x,t) \tau} \xi_j,
\label{rw_gen}
\end{equation}
where $A(x,t)$ is the drift coefficient, representing the average velocity, $B(x,t)$ is the diffusion coefficient, and $\tau$ a (small) time step. The quantity $\xi_j$ is a white noise with zero mean and unit variance. The coefficients $A(x,t)$ and $B(x,t)$ completely define the properties of the microscopic particles dynamics. Remark that Eq.~\eqref{rw_gen} describes a Markovian (memoryless) process: although the evolution of a single trajectory may depend on the others (i.e., the process can in general be nonlinear), knowledge of particles positions at time $t$ is sufficient to determine the displacements at the following step $t+\tau$. It can be shown that the ensemble-averaged density $P(x,t)$ of a particles plume obeying Eq.~\eqref{rw_gen}, i.e.,
\begin{equation}
P(x,t)=\langle \sum_j \delta[x-x_j(t)]\rangle,
\end{equation}
satisfies the (in general nonlinear) Fokker-Planck equation~\cite{risken}
\begin{equation}
\partial_ t P(x,t)=-\partial_x \left[A(x,t) - \partial_x B(x,t) \right]P(x,t).
\label{eq_fp_gen}
\end{equation}
Nonlinearities arise when $A=A(P)$ and/or $B=B(P)$, i.e., when the coefficients depend on particles concentration.

Then, if we want to identify the walkers in Eq.~\eqref{rw_gen} with the microscopic dynamics underlying Eq.~\eqref{eq_ade_gen}, we must properly assign the drift $A(x,t)$ and diffusion $B(x,t)$ of the stochastic process. In other words, we must impose $A(x,t)$ and $B(x,t)$ so that knowledge of $P(x,t)$ provides information on the quantity $p(x,t)$: this in turn establishes a link between the physical variables $\theta(x,t), q(x,t)$, $d(x,t)$ and the parameters $A(x,t)$ and $B(x,t)$.

Letting~\cite{risken}
\begin{equation}
A(x,t)=q(x,t)+\frac{1}{\theta(x,t)}\partial_x\theta(x,t)d(x,t)
\end{equation}
and
\begin{equation}
B(x,t)=d(x,t),
\end{equation}
it is easy to prove that we can identify
\begin{equation}
p(x,t)=P(x,t)/\theta(x,t),
\end{equation}
up to a normalization factor, which provides the desired link.

Let us address Eq.~\eqref{ade_S} first. In this case, the particles that stochastically travel in the porous medium represent fluid parcels that progressively change the saturation distribution in the traversed region~\cite{bhattacharya}. The nonlinearity of the governing equation arises from the fact that the advection and diffusion coefficients depend both on $s(x,t)$. Then, determining the evolution of the saturation profile at time $t+\tau$ requires preliminarily knowing the saturation profile itself at time $t$. In terms of random walks, this implies that particles positions at the following time step can be updated once the positions of all the particles at the current time have been determined. In other words, particles trajectories are correlated via the saturation, as the advection and diffusion coefficients depend on the fluid saturation, $s$. At each time step $\tau$, $s(x,t)$ is first computed on the basis of particles positions at time $t$, then time is updated $t=t+\tau$ and particles are displaced. From the previous considerations (the parameter $\theta$ is assumed to be constant and can be simplified), it follows that
\begin{equation}
x^s_j(t+\tau)=x^s_j(t) + \mu^s + \sigma^s \xi_j,
\label{rw_S}
\end{equation}
where
\begin{equation}
\mu^s=\left[ v(s) + \partial_x \kappa(s) \right] \tau
\label{mu_S}
\end{equation}
and
\begin{equation}
\sigma^s=\sqrt{2 \kappa(s) \tau}.
\label{sig_S}
\end{equation}
In the hydrodynamic limit $\tau \to 0$, the ensemble-averaged fluid parcels profiles $P_s(x,t)$ converge to the solution $s(x,t)$ of Eq.~\eqref{ade_S}.

\begin{figure}[t]
\centerline{\epsfxsize=9.0cm\epsfbox{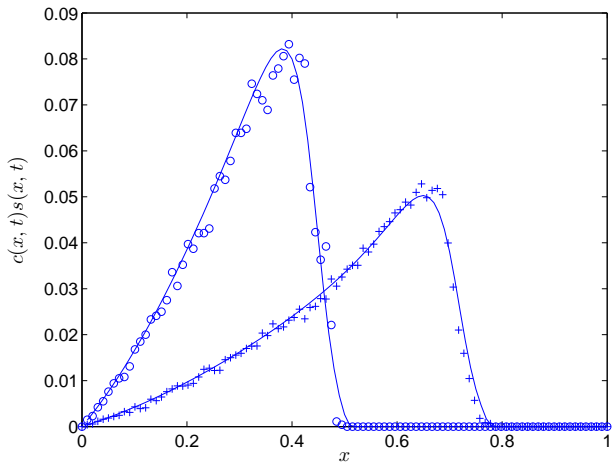}}
\caption{Contaminant concentration profiles $c(x,t)s(x,t)$ at a fixed time $t=1$, for step injection from $t_0=0$ to $t_c=0.1$. The simulation parameters are $\kappa_0=0.05$, $\chi=0.4$, $v_0=0.3$, $\nu=0.2$, $D_0=0.1$, and $\phi=2$. Monte Carlo simulations are shown as symbols (circles at $t=0.5$, crosses at $t=1$), numerical integration as solid lines.}
   \label{fig6}
\end{figure}

Once $s(x,t)$ has been obtained at each time step, the evolution of the concentration profile in Eq.~\eqref{ade_C} can be determined from a second ensemble of particles representing the pollutant parcels. This (linear) random walk must obey
\begin{equation}
x^c_j(t+\tau)=x^c_j(t) + \mu^c + \sigma^c \xi_j,
\label{rw_C}
\end{equation}
where
\begin{equation}
\mu^c=\left[\frac{u(s)}{s}  + \frac{1}{s}\partial_x s D(s) \right] \tau
\label{mu_c}
\end{equation}
and
\begin{equation}
\sigma^c=\sqrt{2 D(s) \tau}.
\label{sig_C}
\end{equation}
Finally, we can identify $c(x,t)=P_c(x,t) / s(x,t)$, where $P_c(x,t)$ is the ensemble-averaged concentration of the contaminant walkers.

In principle, the random walk schemes defined above can be used to determine fluid saturation and contaminant concentration profiles for an arbitrary choice of the time and space dependent coefficients. {In practice, however, because of the sharp gradients and steep profiles resulting from the nonlinearities in Eq.~\eqref{ade_S}, special care is needed in the choice of the numerical values for the time step, $\tau$.}
Also, the evaluation of the space derivatives in the drift terms of the random walk is ill-defined for abrupt jumps (discontinuities) in the equations parameters~\cite{uffink, labolle1, labolle2}. In all such cases, it is convenient to resort to the ad hoc scheme originally proposed in~\cite{labolle1} for particle transport in composite porous media.
For instance, the random walk for the case of fluid parcels would read
\begin{equation}
x^s_j(t+\tau)=x^s_j(t) + v(s) \tau + \sigma^s[x^s_j(t) + \Delta x^s] \xi_j,
\label{lab_S}
\end{equation}
where
\begin{equation}
\Delta x^s =\sigma^s[x^s_j(t)] \xi_j.
\end{equation}
The expression for the case of contaminant particles (with non-constant $\theta$) is more involved and reads
\begin{equation}
x^c_j(t+\tau)=x^c_j(t) + \frac{u(s)}{s} \tau + \frac{1}{\sqrt{s}}\hat{\sigma}^c[x^c_j(t) + \Delta x^c] \xi_j,
\label{lab_C}
\end{equation}
where
\begin{equation}
\hat{\sigma}^c=\sqrt{2 D(s) s \tau}
\end{equation}
and
\begin{equation}
\Delta x^c=\frac{1}{\sqrt{s}} \hat{\sigma}^c [x^s_j(t)] \xi_j.
\end{equation}
In~\cite{labolle1} it has been shown that schemes~\eqref{lab_S} and~\eqref{lab_C} are equivalent to~\eqref{rw_S} and~\eqref{rw_C}, respectively, when coefficients are sufficiently smooth.

\begin{figure}[t]
\centerline{\epsfxsize=9.0cm\epsfbox{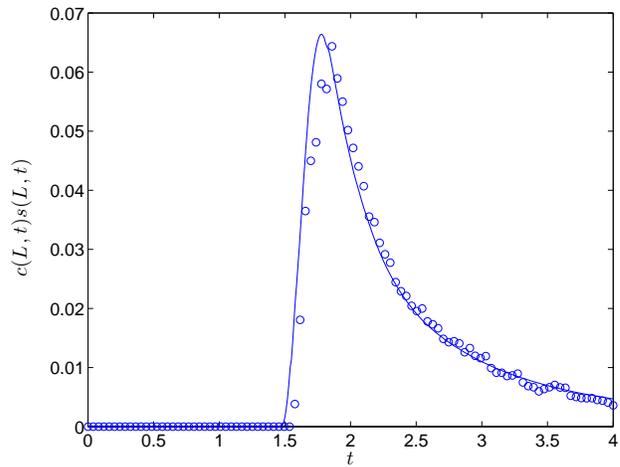}}
\caption{Contaminant profile $c(\ell,t)s(\ell,t)$ measured at the column outlet, as a function of time, for step injection from $t_0=0$ to $t_c=0.1$. The simulation parameters are $\kappa_0=0.05$, $\chi=0.4$, $v_0=0.3$, $\nu=0.2$, $D_0=0.1$, and $\phi=2$. Monte Carlo simulations are shown as symbols, numerical integration as solid line.}
   \label{fig7}
\end{figure}

Finally, just as for the governing equations above, the random walk schemes must be supplemented by the appropriate boundary and initial conditions. For each scheme separately, at time $t=0$ a given (large) number of particles is attributed to each $dx$ along the discretized domain representing the column, so to reproduce the initial saturation distribution and the contaminant concentration profile. A constant saturation level (or concentration) at the inlet is imposed at each time step by keeping the number of particles located at $x=0$ equal to some reference value, i.e., by replacing the walkers that have either come back to the reservoir ($x<0$), or moved towards the interior of the column ($x>0$). The Neumann boundary condition at the outlet is imposed by applying a reflection rule to the diffusive component of the displacement, while particles advected past the outlet during the same time step are removed from simulation. Note that a Dirichlet (absorbing) boundary condition at the outlet would correspond to removing each particle from the ensemble upon touching the column end.

Given an initial particles configuration, these are displaced at each time step according to the rules prescribed above. First, the necessary coefficients are computed at assigned $s(x,t)$ profile (which is known on the basis of flow parcels positions). This allows displacing the walkers at position $x^s_j$ during the time step $\tau$. Then, the $s(x,t)$ profile is updated. Finally, the walkers at position $x^c_j$ are displaced and their profile updated. Up to a normalization factor, the spatial profiles are determined by ensemble-averaging the walkers locations at a fixed time; the breakthrough curves at a given position are determined by counting the net number of walkers crossing that location at each time step. Given the nonlinearities and the coupling between the two schemes, the time step $\tau$ must be chosen sufficiently small to achieve convergence. Moreover, the number of simulated particles must be sufficiently large to attain a good accuracy in the estimated profiles.

\section{Numerical simulations and comparisons}
\label{sim_homogeneous}

We compare now the Monte Carlo simulations of the random walk schemes proposed in Sec.~\ref{random walk} to the numerical solution of the governing equations for $s(x,t)$ and $c(x,t)$. In principle, the random walk schemes are very general and can account for arbitrary functional forms (even discontinuous) of the various coefficients. In the following examples, we will focus on a power-law scaling, which is commonly encountered in the empirical constitutive laws, such as the Van Genuchten or Brooks and Corey laws, and can usually fit experimental data~\cite{bear, bhattacharya, vangenuchten, pachepsky, brooks}. In particular, for the volumetric fluid content we will assume $\kappa(s)=\kappa_0 s^\chi$ and $v(s)=v_0 s^\nu$, $\kappa_0$ and $v_0$ being some constant reference values. The exponents $\chi$ and $\nu$ are material parameters and depend on the details of the microgeometry. Moreover, for the solutes concentration evolution we also assume power-law scaling, $D(s)=D_0 s^\phi$, where $D_0$ is a constant diffusion coefficient and $\phi$ is the scaling exponent.

The physical quantities that we examine in the following are the spatial profiles at a fixed time, which are helpful in estimating the average displacement and spread of the fluid flow and of the contaminant species, and the breakthrough curves at the outlet of the column, which allow assessing the distribution of the times needed to travel from the source to the measure point. In the following, we assume that the column has unit length, $\ell=1$. Furthermore, we assume a constant saturation level $s(0,t)=1$ at the inlet, and let the fluid flow infiltrate the column under the combined action of capillarity and gravity.

Fig.~\ref{fig1} shows the influence of the power-law scaling exponent for the capillary diffusivity, $\chi$, on the spatial profiles of the fluid saturation, $s(x,t)$, at a fixed time $t=1$ ($\chi=0.1, 0.5, 1$ and $2$). Saturation profiles become steeper as $\chi$ increases. Fig.~\ref{fig2} shows the effects of the exponent of the velocity scaling law, $\nu$, on the spatial profiles $s(x,t)$: again, the profiles become steeper as $\nu$ increases, although the variation is milder than for the $\chi$ variation. In both cases, the agreement between Monte Carlo simulation and numerical integration of the corresponding equation is excellent.

In the context of underground contaminant transport, abrupt spatial variations in the physical properties of the traversed media may commonly arise~\cite{uffink, labolle1}. These, in turn, give rise to sharply varying (i.e., possibly discontinuous) transport coefficients and strongly affect particles trajectories. We address one such cases by considering a spatially discontinuous $\kappa_0$: in particular, we assume that at a given interface between two layers $\kappa_0$ has a sudden step variation, while being constant in each layer separately. This situation is usually referred to as a macroscopic heterogeneity (the two layers are thought to be homogeneous at the local scale)~\cite{cortis_gallo}. All the other parameters are constant across the interface. The schemes~\eqref{lab_S} and~\eqref{lab_C} would be suitable to deal also with more involved cases, e.g., multiple heterogeneities. Recent experimental results~\cite{cortis_wrr} suggest that modeling transport through a sharp interface may require skewed flux corrections~\cite{zoia_cortis}. In this work, we do not address this issue and assume that the fluid flux is adequately described by Eq.~\eqref{flux_S}, so that the random walk schemes above hold. In Figs.~\ref{fig3} and~\ref{fig4} we display the breakthrough curves $j_s(\ell,t)$ as a function of time and the spatial saturation profiles $s(x,t)$ at a fixed time, respectively. We compare the curves for the case of fluid flow passing first through the layer at high $\kappa_0=0.4$ and then through the layer at low $\kappa_0=0.05$ with those where flow is in the opposite direction. For the boundary conditions considered here, fluid flow reaches the outlet earlier in the former case (Fig.~\ref{fig3}). Remark that for the chosen boundary and initial conditions the breakthrough curve reaches the saturation value $j_s(\ell,t \to \infty)=v_0$. The saturation profiles along the column show sharp gradients at the interface, while preserving continuity (Fig.~\ref{fig4}). In both cases, the agreement between Monte Carlo simulation and numerical integration of the corresponding equation is excellent.

Note that for the case of non-constant velocity $v$, the breakthrough curve $j_s(\ell,t)=v(s)s \vert_{x=\ell}$ does not coincide (up to a normalization constant) with the saturation $s(\ell,t)$ measured at the outlet. This is immediately apparent from Fig.~\ref{fig5}, where we show Monte Carlo simulations and numerical integrations of the two curves (for the same parameters). This point might be relevant while applying inverse problem techniques to the estimate of model parameters on the basis of experimental data.

\begin{figure}[t]
\centerline{\epsfxsize=9.0cm\epsfbox{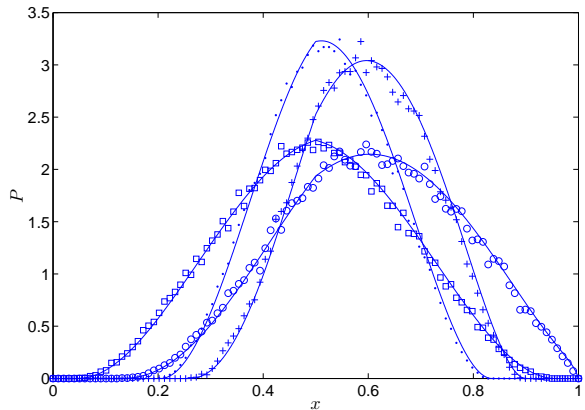}}
\caption{Unsaturated media with trapping processes: fluid flow profiles $P$ at time $t=2.25\, {10}^{-2}$. For all curves, $\alpha=0.6$ and $\lambda=1$. Crosses: $a_0=4$, $a=0.25$, $b_0=0.2$, $b=0.5$. Dots: $a_0=1$, $a=0.25$, $b_0=0.2$, $b=0.5$. Squares: $a_0=0.1$, $a=0.25$, $b_0=0.5$, $b=0.5$. Circles: $a_0=5$, $a=0.25$, $b_0=0.5$, $b=0.5$. The corresponding numerical integration curves are plotted as solid lines.}
   \label{fig8}
\end{figure}

Finally, in Figs.~\ref{fig6} and~\ref{fig7} we display the spatial profiles and the outlet values of the contaminant concentration, respectively, corresponding to a finite-duration step injection. The computed quantity is $c(x,t)s(x,t)$, i.e., the product of concentration and saturation. The spatial profiles are visibly skewed (Fig.~\ref{fig6}), and this behavior is reflected in the long tail of the concentration measured at the outlet of the column as a function of time (Fig.~\ref{fig7}). These features result from the coupling with the saturation and from the nonlinearities involved in the flow-transport processes, and could possibly explain the heavy-tailed breakthrough curves reported in the literature (see, e.g.,~\cite{bromly}) for nonsaturated homogeneous porous media.

\section{Fluid flow through unsaturated heterogeneous media}
\label{heterogeneous}

So far, we have focused our attention on the case of unsaturated homogeneous porous materials, i.e., materials that do not display any significant degree of disorder in the pore-space geometry. The homogeneity hypothesis is mirrored in the Markovian nature of the associated random walks: all spatial sites are statistically equivalent, and particles sojourns have the same duration $\tau$ at each of them, so that trajectories have no memory of past positions. On the other hand, it is well-known that porous media are actually characterized by heterogeneities at multiple scales, which ultimately affect fluid and contaminant particles displacements~\cite{rev_geo, scher_framework}.

\begin{figure}[t]
\centerline{\epsfxsize=9.0cm\epsfbox{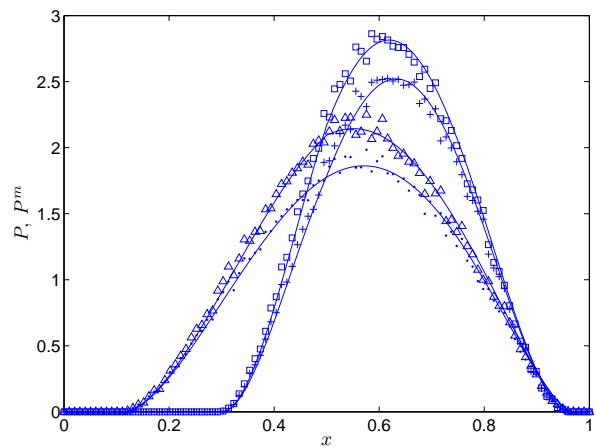}}
\caption{Unsaturated media with trapping processes: fluid flow profiles at time $t=2.25\, {10}^{-2}$. For all curves, $\alpha=0.4$ and $\lambda=1$. Triangles ($P$) and dots ($P^m$): $A=a_0 (P^m)^a$ and $B=b_0 (P^m)^b$, with $a_0=2$, $a=0.4$, $b_0=0.4$, $b=0.6$. Squares ($P$) and crosses ($P^m$): $A=a_0 (P)^a$ and $B=b_0 (P)^b$, with $a_0=6$, $a=0$, $b_0=0.2$, $b=0.5$. The corresponding numerical integration curves are plotted as solid lines.}
   \label{fig9}
\end{figure}

Consider a fluid flow in a complex (possibly fractal) porous microgeometry. In such a situation, the fluid parcels tend to flow in preferential channels~\cite{park}, so that the distribution of the sojourn times at each site is necessarily nonuniform, as suggested by experimental evidence~\cite{kimmich}. A detailed account of Richards' equation inadequacy to explain a number of fluid flow experiments can be found in~\cite{taylor, kuntz, abd, pachepsky} and references therein. In recent years, some extensions of the Richards equation have been proposed to take into account these phenomena~\cite{pachepsky, pachepsky_addendum, gerolymatou}. In particular, it has been conjectured that the wetting front in infiltration processes through nonhomogeneous aggregated media remains immobile for long time periods, and the structural hierarchy of the soil structure is a primary cause of non-Fickian dynamics~\cite{pachepsky, Or}. An expedient means of incorporating such effects into the random walk described by Eq.~\eqref{rw_gen} is to allow for the possibility of trapping events of random duration after each displacement (of duration $\tau$). While several hypotheses can be made on the interplay between displacements and retention times, each corresponding to a distinct conceptual picture of the underlying physical system, here we follow the lines of a continuous time random walk approach called fractal Mobile-Immobile Model (f-MIM)~\cite{Schumer}, which suitably generalizes the discrete-time process defined by Eq.~\eqref{rw_gen}.

Within this framework, it is commonly assumed that trapping times between displacements obey `fat-tailed' (power-law) distributions $\psi_s(t) \sim t^{-1-\alpha}$, $\alpha>0$. If the decay is sufficiently slow, it is not possible to single out a dominant time scale (i.e., the mean of the pdf is not defined), and particles can thus experience a large variation of sojourn times, hence a broad spectrum of effective velocities at each spatial site~\cite{rev_geo}.

More precisely, assume
\begin{equation}
\psi_s(t)= \tau^{1/\alpha}\psi(t/\tau^{1/\alpha})
\label{psis}
\end{equation}
$0<\alpha<1$, $\psi$ being a pdf concentrated on $R^+$, with survival probability $\Psi(t)=\int_t^{+\infty}\psi(t')dt'=\lambda t^{-\alpha}/\Gamma(1-\alpha)+ {\mathcal K}(t)$, ${\mathcal K}$ being integrable. Here $\lambda \ge 0$ is a scaling factor that defines the strength of the trapping events. While in principle $\lambda$ could depend on $x,t$~\cite{zoia_neel}, and even on the particles concentration, here for sake of simplicity we assume that it is constant.

Consider the random walk defined by Eq.~\eqref{rw_gen}. At the end of each displacement, particles wait at the visited spatial site for a random time obeying the pdf $\psi_s(t)$: these sojourn times can be very long as compared to the time scale $\tau$. Because of trapping events, the number of jumps performed by each walker in a given time span may greatly vary: this is an effective means of describing heterogeneous media. We denote by $P^m(x,t)$ and $P^i(x,t)$ the density of mobile and immobile fluid parcels, respectively, and by $P=P^m+P^i$, the total density.

Let $x_{j,n}$ be the position of walker $j$ just after the $n$-th step, which begins at time $t_{j,n}$. This `mobile' step is followed by the $n$-th `immobile' period, with duration $\tau^{1/\alpha}W_n$, where the $W_n$ are independent random variables drawn from $\psi$ (so that the pdf of $\tau^{1/\alpha}W_n$ is $\psi_s$). Then we have
\begin{equation}
x_{j,n+1}=x_{j,n}+ A\tau + \sqrt{2 B\tau} \xi_n,
\label{rwimm_sp}
\end{equation}
and
\begin{equation}
t_{j,n+1}=t_{j,n}+ \tau +\tau^{1/\alpha}W_n.
\label{rwimm_tim}
\end{equation}
As a particular case, when $\lambda=0$ the second equation reduces to $t_{j,n+1}=t_{j,n}+ \tau$ and we recover the homogeneous random walk defined by Eq.~\eqref{rw_gen}. When $A$ and $B$ are uniform and constant, in the hydrodynamic limit ($\tau \to 0$) the particles dynamics above defines a linear fractal-MIM model~\cite{Schumer, Mary}, a generalization of the standard (linear) MIM model~\cite{VieVG}. In this case, it can be shown via subordination that $P$ satisfies an equation akin to Eq.~\eqref{eq_fp_gen}, except that the left-hand side is replaced by $[\partial_t+\lambda \partial_t^\alpha] P$, $\partial_t^\alpha$ being a Caputo derivative of order $\alpha$~\cite{Schumer} (details are provided in Appendix~\ref{A}).

Unsaturated flow in porous media can be characterized by droplets, slug-flows, Darcy's flow, or all of the above. Correspondingly, $A$ and $B$ may depend on $P$ or $P^m$. In highly unsaturated materials, we may conjecture that smaller pores, previously wetted and full of trapped fluid, modify the surface properties of larger pores where mobile fluid flows, so that $A$ and $B$ may depend also on $P^i$. For sake of simplicity, we have neglected here other possibilities: for instance, additional nonlinearities would be introduced at strong solutes concentrations~\cite{zoia_nonlin}. In all such cases, it is more convenient to derive the governing equation for the fluid parcels density by resorting to the relation between $P^m(x,t)$ and $P^i(x,t)$, as in~\cite{Mary}.

We make use of the ancillary pdf $f(x,t)$ for a walker of just being released from a trap at $x$, at time $t$, and denote by $r(x,t)$ a possible source term. Particles that are mobile at time $t$ were either released from a trap, or came from the source at time $t-t'$, with $0<t'<\tau$. We assume that the diffusive step of the displacement occurs at the end of the mobile period (see the discussion in~\cite{zoia_neel}). Moreover, during the time interval $[t-t',t]$, the motion of a walker is determined by the velocity field $A$: we denote the travelled distance by $u_{x,t,t'}$.

Hence, we have
\begin{equation}
P^m(x,t)=\int_0^\tau[f+r](x-u_{x,t,t'},t-t')dt',
\label{Pmfr}
\end{equation}
which implies
\begin{equation}
[f+r](x,u_{x,t,\tau},t-\tau)=\tau^{-1}P_{m}(x,t)+O(\tau)
\label{Pmapprox}
\end{equation}
provided that $A$ (hence $u$) and $f+r$ are smooth~\footnote{This is true when assuming that the term $f+r$ has bounded derivatives (w.r.t. $x$ and $t$), dominated by $t^{-1-\varepsilon}$ with $\varepsilon>0$. This strong assumption implies that applying a time convolution with a bounded kernel will preserve the approximation~\eqref{Pmapprox}. This was not necessary in~\cite{zoia_neel}, where only time-dependent velocity fields were addressed.}.

The density $P^i$ also depends on $f+r$: any immobile particle at time $t$ was trapped at a time $t'<t$, at the end of a mobile step that began at time $t-t'-\tau$ and involved a single diffusive step $\sqrt{2B\tau} \xi$. Denoting by $y$ the amplitude of this latter step, we have
\begin{eqnarray}
& P^i(x,t)=\int_{y\in R} dy \int_0^t dt'\Psi(t'/\tau^{1/\alpha})\varphi_{\sqrt{2B\tau}}(y)\nonumber \\
& [f+r](x-y-u_{x-y,t-t',\tau},t-t'-\tau).
\label{Pifr}
\end{eqnarray}
Here $\Psi(t'/\tau^{1/\alpha})$ represents the probability for a trapping duration to be larger than $t'$, and $\varphi_{\sqrt{2B\tau}}(y) = 1/\sqrt{2B\tau}\varphi(y/\sqrt{2B\tau})$ denotes the pdf of $\xi$, $\varphi$ being the normal distribution. Moreover, $B$ may be nonuniform, and depend on the starting point of each jump ($x-y$ in Eq.~\eqref{Pifr}). In the following, we will assume that $A$ and $B$ depend on $(x,t)$ either directly, or because they are functions of densities such as $P^m$, $P^i$ or $P$.

Denoting time convolutions (in $R^+$) by $*$, i.e., $F*G(t)=\int_0^t F(t-t')G(t')dt'$, and recalling that $\Psi$ is bounded by $1$, we make use of approximation~\eqref{Pmapprox} in Eq.~\eqref{Pifr} and obtain
\begin{eqnarray}
& P^i(x,t)=\tau^{-1}\Psi(t/\tau^{1/\alpha})* \nonumber \\
& \int_{y\in R}P^m(x-y,t)\varphi_{\sqrt{2B(x-y)\tau}}(y)dy+O(\tau).
\label{PiPmfr}
\end{eqnarray}
In the hydrodynamic limit,
\begin{equation}
\int_{y\in R}P^m(x-y,t)\varphi_{\sqrt{2B(x-y)\tau}}(y)dy \to P^m(x,t) \nonumber
\end{equation}
(due to Lemma $4$ of Appendix~\ref{C}), and the (time) convolution of kernel $\tau^{-1}\Psi(t/\tau^{1/\alpha})$ converges to the fractional integral $\lambda I_{0,+}^{1-\alpha}$ (see Appendix~\ref{A}), hence
\begin{equation}
P^i(x,t)=\lambda I_{0,+}^{1-\alpha} P^m(x,t),
\label{Pi}
\end{equation}
which provides a relation between $P^m$ and $P^i$. We can identically rewrite as
\begin{equation}
P(x,t)=[\text{Id}+\lambda I_{0,+}^{1-\alpha}] P^m(x,t).
\label{PPm}
\end{equation}
Then, the inversion of the term $\text{Id}+\lambda I_{0,+}^{1-\alpha}$ yields~\cite{Mary}
\begin{equation}
P^m(x,t)=[\text{Id}+\lambda I_{0,+}^{1-\alpha}]^{-1} P(x,t).
\label{PmP}
\end{equation}
The hydrodynamic limit of the walkers flux ${\mathcal F}(x,t)$ (corresponding to the dynamics in Eqs.~\eqref{rwimm_sp} and~\eqref{rwimm_tim}) is derived in Appendix~\ref{B} and reads
\begin{equation}
{\mathcal F}= [A -\partial_x B ] P^m,
\label{fickfrac}
\end{equation}
This follows from the fact that particles contribute to the flow only when being in the mobile phase $P^m$. Finally, mass conservation $\partial_t P= -\partial_x {\mathcal F}+r$ implies
\begin{equation}
\partial_t P= - \partial_{x} [A - \partial_x B ]P^m + r,
\label{consPPm}
\end{equation}
which is the desired governing equation for transport processes with trapping events~\footnote{Another formulation, equivalent when parameters are constant, is $(\partial_t+\lambda\partial_t^\alpha) P= - \partial_{x}[A - \partial_{x} B]P+(\text{Id}+\lambda I_{0,+}^{1-\alpha})r-\lambda t^{-\alpha}/\Gamma(1-\alpha) P^0 $~\cite{Schumer}.}. Equation~\eqref{consPPm}, rather than being simply postulated as a phenomenological `fractional derivatives generalization' of the standard Richards equation, has been here derived as the hydrodynamic limit of an underlying nonlinear and non-Markovian stochastic process with a definite physical meaning. Remark that Eq.~\eqref{consPPm} is in principle nonlinear, as the coefficients $A$ and $B$ may depend on $P$, $P^m$ or $P^i$, as is the case for unsaturated flows. Moreover, Eq.~\eqref{consPPm} contains a memory kernel (via the relation~\eqref{PPm} between $P^m$ and $P$): in other words, because of the slowly-decaying time kernel, knowledge of the past history of the particle is required in order to determine its future displacement. Therefore, the fluid parcels density is affected at the same time by nonlinearities and memory effects: the relative strength of these components ultimately determines the fate of the flow in the traversed media.

Observe moreover that Eq.~\eqref{consPPm} is written in `Fokker-Planck form', i.e., with a term of the kind $\partial_{x}\partial_x B P^m$. If we were to postulate a conservative (Fickian) formulation for the flux, corrections to the drift coefficient $A$ should be introduced, in analogy with the case of the standard Richards equation for homogeneous media.

The behavior of fluid flow with nonlinear coefficients and retention times is illustrated in Figs.~\ref{fig8} and~\ref{fig9}. In particular, we proceed to compare the Monte Carlo simulations of the random walks described by Eqs.~\eqref{rwimm_tim} and~\eqref{rwimm_sp} (in the hydrodynamic limit) with the numerical integration of the governing Eqs.~\eqref{Pi} and~\eqref{consPPm}. Monte Carlo simulations proceed along the same lines as for homogeneous media. Concerning the numerical integration, we found more expedient to recast Eq.~\eqref{consPPm} in the equivalent formulation
\begin{equation}
[\partial_t+\lambda D^\alpha_t] P^m =-\partial_x [A -\partial_x B]P^m+r,
\label{RLconsPm}
\end{equation}
$D^\alpha_t$ being a Riemann-Liouville fractional derivative, defined in Appendix~\ref{A}. Once Eq.~\eqref{RLconsPm} has been solved for $P^m$, $P$ is easily computed from Eq.~\eqref{PPm}. We discretized Eq.~\eqref{RLconsPm} according to a semi-implicit scheme as in~\cite{Mary}, so to avoid possible instabilities connected with nonlinearities.

In Fig.~\ref{fig8} we display the total fluid flow density $P$ at a given time, for a different values of the coefficients. We make the hypothesis that $A$ and $B$ depend on the mobile phase $P^m$, with a power-law scaling $A=a_0 (P^m)^a$ and $B=b_0 (P^m)^b$ (similarly as done for the homogeneous media). The initial condition is a fluid pulse located at $x_0=\ell/2$. Absorbing boundary conditions are set at either end of the medium. In Fig.~\ref{fig9} we display the total fluid flow density $P$ as compared to the mobile density $P^m$, when the parameters $A$ and $B$ separately depend (with a power-law scaling) on the mobile or total fluid density. Boundary and initial conditions are the same as in the previous example. Both figures show a very good agreement between Monte Carlo simulation and numerical integration.

Finally, replacing Eq.~\eqref{psis} with $\psi_s(t)=\tau\psi(t/\tau)$ yields $P^i=\lambda P^m$ when $\psi$ has a finite average $\lambda$, i.e., when the pdf decays sufficiently fast at long times. In this case, Eq.~\eqref{consPPm} holds with $P^m=(1+\lambda)^{-1}P$, and we have
\begin{equation}
(1+\lambda)\partial_t P =-\partial_x [A -\partial_x B]P+(1+\lambda)r,
\label{RLclassc}
\end{equation}
which is a nonlinear Fokker-Planck equation with a retardation factor $\lambda$~\cite{zoia_neel}.

\section{Discussion and conclusions}
\label{conclusions}

In this work we have addressed nonlinear coupled flow and transport processes in variably saturated porous media. After briefly recalling the governing equations for homogeneous materials, we have shown how to describe fluid and solutes parcels trajectories by resorting to a unified random walk framework. Monte Carlo simulations of the underlying microscopic particles dynamics have been successfully compared to the numerical solutions of the governing equations. The random walk approach has turned out to be a flexible tool, which allows dealing with nonlinear and/or discontinuous transport coefficients.

Then, we have introduced a nonlinear f-MIM continuous time random walk scheme aimed at describing fluid flow through variably saturated heterogeneous media. The effects of spatial heterogeneities are mirrored in the possibility of long (power-law distributed) sojourn times of the flowing particles at each visited site. The corresponding governing equations have been derived and their numerical integration has been then compared with the Monte Carlo simulations of the walkers dynamics. We remark that the proposed f-MIM generalization of nonlinear transport equations is not unique in any respect. Indeed, other possible approaches have been illustrated, e.g., in~\cite{sanchez1, sanchez2} and in~\cite{Lutsko} by means of a generalized Montroll-Weiss master equation with a jump pdf depending on the walkers density.

Although focus has been given to fluid flow through heterogeneous unsaturated materials, the proposed nonlinear f-MIM scheme is fairly general and can be straightforwardly applied to the coupled solutes transport problem, as well. Similarly as fluid parcels can be affected by the irregular geometry of the traversed material, the solutes concentration can also experience the effects of nonhomogeneities, due, e.g., to chemical-physical exchanges of the solute species with the surrounding environment~\cite{berkowitz_sorp}. In analogy with the case of fluid flow dynamics, we may then take into account these contributions by introducing a power-law distribution $\psi_c(t) \sim t^{-1-\beta}$, with $\beta>0$, for the waiting times of contaminant particles between displacements. This pdf characterizes the sorption times of the transported species within the medium. In general, there is no reason to suppose that the exponent $\beta$ coincides with $\alpha$. A similar distinction between flow- and solutes-induced retention times has been previously introduced in~\cite{rev_geo, berkowitz_sorp, cortis_crypto} within a (linear) CTRW framework.

In summary, the solutes concentration in heterogeneous variably saturated materials may display deviations from standard Fickian behavior due to three concurrent processes: $i)$ the space- and time-varying saturation profile within the medium (nonlinear effects), $ii)$ the spatial heterogeneities experienced by the fluid flow (memory effects), and $iii)$ the spatial heterogeneities experienced by the solutes (memory effects). As experimental data such as contaminant profiles or breakthrough curves are usually limited and/or affected by measurement noise, distinguishing the effects of spatial heterogeneities on flow and transport processes separately is an highly demanding task, and many research efforts have been recently devoted to this aim (see, e.g.,~\cite{berkowitz_sorp} and references therein). At present, it is therefore an open question whether these distinct contributions could be separately analyzed on the basis of measured data, or rather described by an effective CTRW model: we will discuss in detail this topic in a forthcoming paper.

As a final remark, note that in this work we have confined our attention to the migration of nonreacting (passive) species. Nonetheless, all the random walk algorithms introduced here could be easily extended so as to describe the transport of radionuclides, by computing the decay time before simulating the particle trajectory~\cite{zoia_reactive}.

\appendix

\section{Fractional integrals and derivatives}
\label{A}

The fractional integral $I_{0,+}^{\alpha} f$ of order $\alpha >0$ is
\begin{equation}
I_{0,+}^{\alpha}f(t)=\frac{1}{\Gamma(\alpha)}\int_0^t(t-t')^{\alpha-1}f(t')dt', \nonumber
\end{equation}
which is a generalization of the usual multiple integrals to arbitrary (positive) order~\cite{Rub, Sa}.

The Caputo fractional derivative $\partial_t^\alpha f$ of order $n<\alpha<n+1$ is 
\begin{equation}
\partial_t^\alpha f(t)=I_{0,+}^{n+1-\alpha}\partial^{n+1}_t f(t), \nonumber
\end{equation}
$n$ being an integer~\cite{Sa, MaiUd, Kilb}. The Riemann-Liouville fractional derivative $D_t^\alpha$ is instead defined as
\begin{equation}
D_t^\alpha f(t)=\partial^{n+1}_tI_{0,+}^{n+1-\alpha} f(t) \nonumber
\end{equation}
for $n<\alpha<n+1$. Note that the Riemann-Liouville derivative applies to slightly more general functions than Caputo's, and when both can be applied (i.e., for differentiable functions $f$) we have $D_t^\alpha f(t)=\partial_t^\alpha f(t)+t^{-\alpha}f(0+)/\Gamma(1-\alpha)$ for $0<\alpha<1$.

In Section~\ref{heterogeneous} we use the following Lemma.\\

{\em Lemma 1.} {\it(i)} Suppose $\Psi$ is a positive-valued, decreasing function defined on $R^+$, with $\Psi(0)=1$ and $\Psi(t)=t^{-\alpha} /\Gamma(1-\alpha)+{\mathcal K}(t)$, with ${\mathcal K}$ integrable and $0<\alpha<1$. Then, the convolution of kernel $\tau^{-1} \Psi(t/\tau^{1/\alpha})$ converges to $I_{0,+}^{\alpha} $ in $L^p(R^+)$, with $1\leq p\leq+\infty$, when $\tau\to 0$.\\

{\it(ii)} If $\Psi$ is integrable, then the convolution of kernel $\tau^{-1}\Psi(t\tau^{-1})$ converges to $\int_{R^+}\Psi(t) dt \text{Id}$.\\

{\em Proof.} Due to the above definitions, the convolution of kernel $\tau\Gamma(1-\alpha)^{-1}(t/\tau^{1/\alpha})^{-\alpha}$ is exactly $I_{0,+}^{1-\alpha}$. Moreover, the convolution of kernel $\tau^{-1/\alpha}{\mathcal K}(t/\tau^{1/\alpha})$ (as a mapping of $L^p(R^+)$) is an approximation to $\int_{R^+}{\mathcal K}(t)dt\text{Id}$~\cite{Sa, Rub}, so that the convolution of kernel $\tau^{-1}{\mathcal K}(t/\tau^{1/\alpha})$ tends to zero when $\tau\to 0$ (due to $\alpha<1$), which proves point {\it(i)}. Point {\it(ii)} is a direct consequence of the reference~\cite{Sa, Rub} concerning approximations to $\text{Id}$.\\

{\em Remark.} With $\varphi_\ell(y)=1/\ell\varphi(y/\ell)$, the space convolution (denoted  by $\star$, with, $f\star g(x)=\int_Rf(x-x')g(x')dx'$) of kernel $\varphi_\ell$ converges to $\text{Id}$ when $\ell \to 0$~\cite{Sa, Rub}, in $L^p$, and we have $\varphi_\ell\star G(x)\to G(x)$ pointwise when $G$ is a  continuous function.

\section{Probability current}
\label{B}

We would like to prove Eq.~\eqref{fickfrac}. Let us first introduce the probability current of the random walk defined by Eqs.~\eqref{rwimm_tim} and~\eqref{rwimm_sp}. The current is the probability for a walker to cross a given point $x$ to the right during a time interval $dt$, minus the probability of crossing to the left, divided by $dt$. Upon multiplication by the total number of walkers involved in the random walk, it yields the average number of particles that cross $x$ per unit time.

Note that a walker that is not at the end of the current mobile period crosses $x$ during $[t-dt,t]$, provided that it was released from the trap at time $t-t'$ (with $0<t'<\tau$), between $x-u_{x,t,t'}-A(x-u_{x,t,t'},t-t')dt$ and $x-u_{x,t,t'}$, if $A>0$. This occurs with probability
\begin{eqnarray}
&\int_0^\tau[f+r](x-u_{x,t,t'},t-t') A(x-u_{x,t,t'},t-t')dt dt'\nonumber \\
&\simeq P^m(x,t) A(x,t)dt\nonumber
\end{eqnarray}
when $\tau \to 0$. Moreover, a walker crosses $x$ by a diffusive step with probability $\int_{y>0}[f+r](x-y-u_{x,t,\tau},t-\tau)\Phi(y/\sqrt{2\tau B(x-y)})dy$ per unit time, where $\Phi(z)=\int_z^{+\infty}\varphi(y)dy$. The quantity $\Phi(y/\sqrt{2\tau B})$ represents the probability for a given step to be larger than $y$. This latter probability approximates $\tau^{-1}\int_{y>0}P^m(x-y,t)\Phi(y/\sqrt{2\tau B(x-y)})dy$ when $\tau \to 0$, due to Eq.~\eqref{Pmapprox}. Collecting finally contributions of jumps to the left, the probability current of the random walk in Eqs.~\eqref{rwimm_tim} and~\eqref{rwimm_sp} is
\begin{eqnarray}
& P^m A(x,t)+\tau^{-1}\int_{y>0}P^m(x-y,t)\Phi(y/\sqrt{2\tau B(x-y)}-\nonumber \\
& P^m(x+y,t)\Phi(y/\sqrt{2\tau B(x+y)})dy.\nonumber
\end{eqnarray}
Appendix~\ref{C} shows that the above integral converges to $-\partial_x B P^m \int_{R}y^2\varphi(y)dy$, due to the rapid decrease at infinity of $\varphi$ and $\Phi$. Hence, in the hydrodynamic limit the probability current is given by Eq.~\eqref{fickfrac}.

\section{Technical lemma}
\label{C}

{\em Lemma 2.} Let $\Phi$ be a differentiable function, integrable over $R^+$, positive and decreasing. Suppose also that $B$, as a function of $x$, has a derivative $B'$ satisfying $|yB'(x+y)/B(x+y)|<2$ everywhere. Suppose then that $G$ is an integrable function whose derivative is uniformly bounded, and set $I_{\pm}=\tau^{-1}\int_{y>0}G(x\pm y)\Phi(y/\sqrt{2\tau B(x\pm y)})dy$. Then, the quantity $-I_++I_-$ converges to $-4\partial_x(G(x)B(x))\int_0^{+\infty}z\Phi(z)dz$ when $\tau\to 0$.\\

{\em Remark.} We have $-4\int_0^{+\infty}z\Phi(z)dz=\int_Rz^2\varphi(z)dz$, given that $\varphi(z)=-\Phi'(z)$.\\

{\em Proof of Lemma 2.} We first show that {\it(i)} $-I_++I_-$ is of the form $2\varepsilon^{-1}\int_0^{+\infty}F(\varepsilon z)\Phi(z)dz$, $F$ being a derivable function satisfying $F(0)=0$. Then, {\it(ii)} we apply Lemma $3$ further below.

In view of {\it(i)}, let us fix $x$. The hypotheses of the Lemma allow for a change of variables $z=g_\pm(y)/\sqrt{2\tau}$ in $I_\pm$, with $g_\pm(y)=y/\sqrt{B(x\pm y)}$. The inverse of $g_\pm$ is $h_\pm$, with
\begin{eqnarray}
& g'_\pm(y)=1/\sqrt{B(x\pm y)}[1\mp yB'(x\pm y)/(2{B(x\pm y)})],\nonumber \\
& h'_\pm(Z)=\sqrt{B(x\pm h_\pm( Z))}[1\mp\frac{h_\pm(Z)B'(x\pm h_\pm(Z))}{2B(x\pm h_\pm(Z))}]^{-1},\nonumber
\end{eqnarray}
and $h_\pm(0)=0$. With these notations, set $\varepsilon=\sqrt{2\tau}$ and
\begin{equation}
F(Z)=G(x-h_-(Z))h'_-(Z)-G(x+h_+(Z))h'_+(Z)
\end{equation}
so that letting $y=h_\pm(\varepsilon z)$ in $I_\pm$ yields $-I_++I_-=2\varepsilon^{-1}\int_0^{+\infty}F(\varepsilon z)\Phi(z)dz$.

Now, to apply Lemma $3$, note that $F'(0)=-2G'(x)B(x)+G(x)(h''_-(0)-h''_+(0))$, and that $h''_\pm(Z)=a_\pm\pm b_\pm$, with
\begin{equation}
a_\pm=\pm\frac{h'_\pm(Z)B(x\pm h_\pm(Z))}{2\sqrt{B(x\pm h_\pm(Z))}}[1\mp\frac{h_\pm(Z)B'(x\pm h_\pm(Z))}{2B(x\pm h_\pm(Z))}]^{-1} \nonumber
\end{equation}
and
\begin{equation}
b_\pm=\frac{\sqrt{B(x\pm h_\pm(Z))}}{[1\mp\frac{h_\pm(Z)B'(x\pm h_\pm(Z))}{2B(x\pm h_\pm(Z))}]^2}\frac{c_\pm}{2B(x\pm h_\pm(Z))^2}, \nonumber
\end{equation}
with
\begin{eqnarray}
& c_\pm=h'_\pm(Z)BB'(x\pm h_\pm(Z))\pm h_\pm h'_\pm(Z)BB''(x\pm h_\pm(Z)) \nonumber \\
& \mp h_\pm(Z)h'_\pm(Z)B'^2(x\pm h_\pm(Z)),\nonumber
\end{eqnarray}
which implies $h''_\pm(0)=\pm B'(x)$, so that $F'(0)=-2(GB)'(x)$. Hence, Lemma $2$ is a consequence of the Lemma $3$ below, which itself is included in Lemma $3$ of~\cite{zoia_neel}.

{\em Lemma 3.} Let $\Phi$ be a differentiable function, integrable over $R^+$, and bounded. Then, for any integrable function $F$ satisfying  $F(0)=0$ and whose derivative is uniformly bounded, the expression
\begin{equation}
\varepsilon^{-1}\int_0^{+\infty}F(\varepsilon z)\Phi(z)dz\nonumber
\end{equation}
converges to $F'(0)\int_0^{+\infty}z\Phi(z)dz$ when $\varepsilon\to 0$.\\

In Section~\ref{heterogeneous}, we use the following statement.\\

{\em Lemma 4.} Let $G$ be a continuous bounded function, with $ \varphi$ a pdf. Suppose also that $B$ is as in Lemma $2$. Then, $I=\int_{R}G(x- y)\varphi(y/\sqrt{2\tau B(x- y)}) dy/\sqrt{2\tau B(x- y)} \to G(x)$ when $\tau\to 0$.\\

As in the proof of Lemma $2$, the change of variables $g(y)=y/\sqrt{B(x- y)}$ has an inverse, which we denote by $h$. Thus, we have $I=\int_{R}G(x- h(\sqrt{2\tau}z))\varphi(z)dz$. Hence, the remark of Appendix~\ref{A} proves the statement.

\acknowledgments

A.~C.~was supported by a grant (Order No.~7220) from the Swiss National Cooperative for the Disposal of Radioactive Waste (Nagra), Wettingen, Switzerland: the support was provided to Berkeley Lab through the U.S. Department of Energy Contract No.~{DE-AC02-05CH11231}. We thank Stefan Finsterle and Philippe Montarnal for interesting scientific discussions.

\end{document}